\documentclass[9pt,twocolumn,twoside]{osajnl}

\journal{josab} 

\setboolean{shortarticle}{false}
\usepackage{braket}

\def\bcen{\begin{center}}
\def\ecen{\end{center}}
\usepackage{epstopdf}
\ifthenelse{\boolean{shortarticle}}{\colorlet{color2}{color2b}}{\colorlet{color2}{color2}} 

\title{Intrinsic degree of coherence of two-qubit states and measures of two-particle quantum correlations}

\author[1,*]{Nilakantha Meher}
\author[1,**]{Abu Saleh Musa Patoary}
\author[1,++]{Girish Kulkarni}
\author[1,$\dagger$]{Anand K. Jha}%

\affil[1]{Department of Physics, Indian Institute of Technology Kanpur, Kanpur, UP 208016, India.}
\affil[**]{Present address: Department of Physics, University of Maryland, College Park, Maryland, 20742, USA}
\affil[++]{Present address: Department of Physics, University of Ottawa, Ottawa, Ontario K1N6N5, Canada}
\affil[*]{nilakantha.meher6@gmail.com}
\affil[$\dagger$]{akjha9@gmail.com}

\dates{Compiled \today}

\ociscodes{ (270.0270) Quantum Optics (030.1640)   Coherence}

\doi{\url{http://dx.doi.org/10.1364/ao.XX.XXXXXX}}

\begin{abstract}
Recently, a basis-invariant measure of coherence known as the intrinsic degree of coherence has been established for classical and single-particle quantum states [JOSA B {\bf 36}, 2765 (2019)]. Using the same mathematical construction, in this article, we define the intrinsic degree of coherence of two-qubit states  and demonstrate its usefulness in quantifying two-particle quantum correlations and entanglement. In this context, first of all, we show that the intrinsic degree of coherence of a two-qubit state puts an upper bound on the violations of Bell inequalities that can be achieved with the state and that a two-qubit state with intrinsic degree of coherence less than $1/\sqrt{3}$ cannot violate Bell inequalities. We then show that the quantum discord of a two-qubit state, which quantifies the amount of quantum correlations available in the two-qubit state for certain tasks, is bounded from above by the intrinsic degree of coherence of the state. Next, in the context of two-particle entanglement, we show that the range of values that the concurrence of a two-qubit state can take is decided by the intrinsic degree of coherence of the two-qubit state together with that of the individual qubits. Finally, for the polarization two-qubit states generated by parametric down-conversion of a pump photon, we propose an experimental scheme for measuring the intrinsic degree of coherence of two-qubit states. We also present our  theoretical study showing how the intrinsic degree of coherence of a pump photon dictates the maximum intrinsic degree of coherence of the generated two-qubit state. 
\end{abstract}

\setboolean{displaycopyright}{true}

\begin{document}

\maketitle
\thispagestyle{fancy}

\ifthenelse{\boolean{shortarticle}}{\ifthenelse{\boolean{singlecolumn}}{\abscontentformatted}{\abscontent}}{}
\section{Introduction}

Coherence is the essential ingredient that drives many classical  \cite{redding2012natphot, karamata2004optlett} and quantum technologies \cite{Gisin, Giovannetti, Nielsen, Gisin2}, and in the last several decades quantification of coherence has been a subject of intense research investigations \cite{Mandel,Glauber, Sudarshan1, Wolf}. In the context of a partially polarized field represented by a $2\times 2$ polarization matrix $\rho$, coherence of the state is quantified in a basis-invariant manner using the so-called degree of polarization $P_2$, defined as $P_2=\sqrt{2 \text{Tr}(\rho^2)-1}$ \cite{Mandel}. Since the trace of a matrix is basis-independent, $P_2$ remains invariant under unitary transformations. Although $P_2$ was intended to quantify the coherence of a two-dimensional polarization matrix, it can be used as a basis-independent quantifier of coherence for any two-dimensional classical or quantum state, such as a two-level atomic states or a spin-1/2 particle states. This basis-independent way of defining coherence, which in the context of two-dimensional polarization states yields the degree of polarization, has been extended not only to $N$-dimensional states \cite{Barakat, Yao, Patoary} but also to infinite-dimensional states \cite{Patoary}. Such a quantifier is now being referred to as the \textquotedblleft intrinsic degree of coherence\textquotedblright \cite{Patoary}.

In the context of entangled two-particle systems, it is known that the system possesses two-particle coherence in addition to one-particle coherences possessed by the individual subsystems. This two-particle coherence is responsible for two-particle interference produced by such systems \cite{Fearn}. Some of the very important examples of two-particle interference include Hong-Ou-Mandel effect \cite{Hong}, Franson interferometer \cite{franson}, frustrated two-photon creation \cite{Herzog}, induced coherence without induced emission \cite{Zou}, and violations of Bell inequality \cite{ou1988prl, shih1988prl}. Among the states that produce two-photon interference, the two-qubit states have been  studied the most. A two-qubit state consists of two subsystems, each of which exists in a two-dimensional Hilbert-space. Our first aim in this article is to quantify the two-particle coherence of two-qubit states in a basis-independent manner. We do so by defining the intrinsic degree of coherence of two-qubit states using the same mathematical construction \cite{Barakat, Yao, Patoary} that has been used for defining the intrinsic degree of coherence of classical and one-particle quantum states.

The two-qubit states are not only the necessary ingredients of many quantum information-based applications \cite{ekert1991prl, bennett1992prl, bennett1993prl} but also used for demonstrating violations of Bell inequalities and thereby for ruling out any potential hidden variable interpretation of quantum mechanics \cite{ou1988prl, shih1988prl}. In the last several years, much effort has gone into quantifying the entanglement of two-qubit states, and among the available entanglement quantifiers, Wootters’s concurrence \cite{wootters1998prl} is the most widely used one. More recently, there also has been an interest in quantifying quantum correlations of two-qubit states. Such correlations are not captured completely by an entanglement measure. One such quantifier is discord, which quantifies the amount of quantum correlations available in the state for certain tasks. As our next aim in this article, we establish important connections between the intrinsic degree of coherence of a two-qubit state and several above-mentioned measures of two-qubit quantum correlation and quantum entanglement. This way, we demonstrate the usefulness of the intrinsic degree of coherence of two-qubit states in quantifying two-particle correlations and entanglement. Finally, for the polarization two-qubit states generated by parametric down-conversion of a pump photon, we propose an experimental scheme for measuring the intrinsic degree of coherence of the two-qubit state. We also present our theoretical study showing how the intrinsic degree of coherence of a pump photon dictates the maximum intrinsic degree of coherence of the generated two-qubit state. 

This article is organized as follows. In Sec. \ref{P22twoqubit}, we define the intrinsic degree of coherence of a two-qubit state. In Sec.~\ref{Correlation}, we establish how it is connected with various measures of quantum correlation, namely, the degree of Bell-violation and the discord. In Sec. \ref{Concurrence}, we discuss how it is related to the concurrence. In the context of polarization two-qubit states produced by the parametric down-conversion of a pump photon, we propose an experimental scheme for measuring the intrinsic degree of coherence of two-qubit states in Sec.~\ref{Experiment} and we discuss how the intrinsic degree of coherence of the pump photon affects the intrinsic degree of coherence of the down-converted two-qubit state in Sec.~\ref{SPDCDiscord}. Finally, we summarize our results in Sec. \ref{Summary}.

\section{Defining the intrinsic Degree of coherence of two-qubit state}\label{P22twoqubit}

Let us consider a  general two-particle system consisting of two subsystems $A$ and $B$ with the Hilbert space dimension of each subsystem being equal to 2. Let us represent this two-qubit state by a $4\times 4$ density matrix $\rho$. We assume the two-qubit state $\rho$ to be normalized, that is, $\text{Tr}(\rho)=1$.  Following the mathematical construction used in Ref. \cite{Patoary}, we define the intrinsic degree of coherence $P_{2\otimes 2}$ of the two-qubit state $\rho$ to be
\begin{align}\label{IDOC}
P_{2\otimes 2}=\sqrt{\frac{4 \text{Tr}(\rho^2)-1}{3}}.
\end{align}
The intrinsic degree of coherence of subsystem $A$ can be written as \cite{Luis,Yao,Patoary}
\begin{align}\label{IDoCSubSystem}
P_2^A=\sqrt{2 \text{Tr}(\rho_A^2)-1},
\end{align} 
where $\rho_A=\text{Tr}_B(\rho)$ is the reduced density matrix of subsystem $A$ and $\text{Tr}_B$ represents the partial trace over subsystem $B$. Similarly, one can write the intrinsic degree of coherence of subsystem $B$ by taking the partial trace over subsystem $A$. Here, we have used the following convention for denoting the intrinsic degree of coherences. For a one-particle system, we simply write the dimensionality of the system as the subscript. For a two-particle system, the subscript consists of two numbers separated by $\otimes$ sign, with the numbers being equal to the dimensionalities of the individual subsystems. We note that  $P_{2\otimes 2}$ is invariant under global unitary transformation and ranges from 0 to 1, that is, $0\leq P_{2\otimes 2}\leq 1$. $P_{2\otimes 2}$ is unity for a pure two-qubit state and is zero for a completely mixed two-qubit state. We further note that just as $P_2$ is called the degree of polarization in the context of single-particle polarization states, $P_{2\otimes 2}$ can analogously be referred to as the degree of two-photon polarization in the context of polarization two-qubit states. In general, $P_2^A \neq P_2^B$; however, when $\rho$ is a pure two-qubit state, $P_2^A=P_2^B$.

\section{Quantum correlation measures and intrinsic degree of coherence}\label{Correlation}

\subsection{Non-local correlation and intrinsic degree of coherence}\label{nonlocal}

Nonlocality is an intriguing feature of quantum-entangled systems and has been established by the experimental demonstrations of the violations of Bell inequalities \cite{Bell, Clauser}. The CHSH form of Bell's inequality involves construction of a Bell parameter $S$ \cite{Clauser}. If for a given state $\rho$ there exists a measurement setting for which $S>2$ then the Bell's inequality is said to be violated and the state $\rho$ is said to be quantum-entangled in the sense that a local hidden variable description of correlations exhibited by the state becomes impossible \cite{Cirelson}.

We now ask the following question: how is the intrinsic degree of coherence $P_{2\otimes 2}$ of a two-qubit state $\rho$ related to the maximum Bell violation realizable with the state? In order to answer this question, first we note that for a given state $\rho$ the Bell parameter can be written in terms of the participation ratio $R=1/\text{Tr}(\rho^2)$ as \cite{Batle}
\begin{align}
S &\leq \sqrt{8/R} \quad {\rm for}  \quad R \in [1,2], \\
S &\leq 2\sqrt{4/R-1} \quad {\rm for} \quad R \in [2,4].
\end{align}
Using the definition given in Eq.~(\ref{IDOC}), we write the above inequalities as
\begin{align}
S &\leq \sqrt{6 P_{2\otimes 2}^2+2} \quad {\rm for} \quad P_{2\otimes 2} \in \left[\frac{1}{\sqrt{3}},1 \right],\\
S &\leq 2\sqrt{3} P_{2\otimes 2} \quad {\rm for} \quad P_{2\otimes 2} \in \left[0, \frac{1}{\sqrt{3}} \right]. 
\end{align}
We note that only the pure two-qubit states with $P_{2\otimes 2}=1$ can exhibit maximum Bell violation with $S=2\sqrt{2}$ and that any state having $P_{2\otimes 2}\leq 1/\sqrt{3}$ cannot violate Bell's inequality. Thus we find that the intrinsic degree of coherence of a two-qubit state puts an upper bound on the maximum Bell violation achievable with the state over all possible measurement settings.

\subsection{Quantum discord and intrinsic degree of coherence of two-qubit states}\label{Discord}

Quantum discord is a measure of quantum correlation \cite{Ollivier}. The evaluation of quantum discord for an arbitrary two-qubit state is a challenging task since that requires optimization of the conditional entropy between the two subsystems \cite{Ollivier}. Closed form analytic expressions for discord are known only for certain classes of states \cite{Adesso,Luo,Chen}. The difficulty in calculating quantum discord led to the introduction of the geometric discord $D_G$ \cite{Dakic}, which for a state $\rho$ is defined as the minimum distance between the given state and the set of zero discord states, that is,
\begin{align}\label{geometricdiscord}
\sqrt{D_G}=\text{min}_{\chi \in \Omega} ||\rho-\chi||,
\end{align}
where $||X||=\sqrt{\text{Tr}(X^\dagger X)}$ is the Frobenius norm. Here $\Omega$ is the set of zero discord states. The elements of $\Omega$ are density matrices of the form \cite{Luo2}
\begin{align}
\chi=\sum_{i} p_i \ket{i}\bra{i}\otimes \rho_{Bi},
\end{align}  
where $\{p_i\}$ are the probabilities, $\{\rho_{Bi}\}$ is the set of all possible states from subsystem $B$ and $\{\ket{i}\}$ is an orthonormal set of basis vectors corresponding to subsystem $A$.

We now show that the intrinsic degree of coherence of a two-qubit state decides the upper bound for quantum discord of the state. To this end, first of all, using the triangle inequality, we write Eq. (\ref{geometricdiscord}) as
\begin{align}\label{traingleinequality}
\sqrt{D_G}&=\text{min}_{\chi \in \Omega} \left|\left|\rho-\frac{I}{4}+\frac{I}{4}-\chi\right|\right|,\nonumber\\
&\leq  \left|\left|\rho-\frac{I}{4}\right|\right|+\text{min}_{\chi \in \Omega}\left|\left|\frac{I}{4}-\chi\right|\right|.
\end{align} 
Since the geometric discord of state $I/4$ is zero, we write Eq. (\ref{traingleinequality}) as
\begin{align}
\sqrt{D_G}\leq  \left|\left|\rho-\frac{I}{4}\right|\right|.
\end{align}
One of the interpretations of $P_{2\otimes 2}$ is that it is the Frobenius distance between $\rho$ and the maximally incoherent state $I/4$, that is, $P_{2\otimes 2}=\sqrt{\frac{4}{3}}\left|\left|\rho-\frac{I}{4}\right|\right|$ \cite{Yao, Patoary}. Using this result, we write the above inequality as
\begin{align}
\sqrt{D_G}\leq  \sqrt{\frac{3}{4}}P_{2\otimes 2}. \label{geometric}
\end{align}
Furthermore, it is known that $D\leq \sqrt{2D_G}$ \cite{Girolami}. Therefore, we write Eq.~(\ref{geometric}) as
\begin{align}\label{DiscordAndP22}
D\leq \sqrt{\frac{3}{2}}P_{2\otimes 2}.
\end{align}
Thus, we find that the quantum correlations present in a two-qubit state as quantified by discord follows an upper bound decided by $P_{2\otimes 2}$ of the two-qubit state. We note that mixed states with $P_{2\otimes 2}<\sqrt{2/3}$ can not have unit quantum discord and that states with $P_{2\otimes 2}=0$, that is, the maximally incoherent states can not have non-zero quantum discord. Because of the unavailability of an analytic expression of quantum discord for a generic two-qubit state, it is very difficult to verify whether or not the bound given by the inequality in Eq.~(\ref{DiscordAndP22}) is saturable. Nevertheless, since $P_{2\otimes 2}$ can be calculated for any given two-qubit state, Eq.~(\ref{DiscordAndP22}) provides an upper bound on discord even in situation in which computing it is non-trivial.

\begin{figure}
\centering
\includegraphics[width=8cm,height=5.5cm]{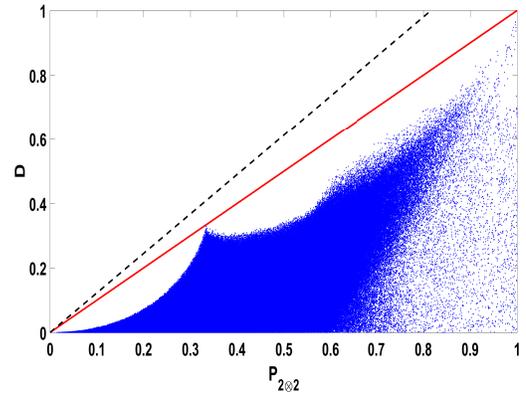}
\caption{ Scatter plot of quantum discord $D$ and $P_{2\otimes 2}$ of $X$-states. We have taken 2$\times 10^7$ number of randomly chosen $X$-states. The continuous line represents $D=P_{2\otimes 2}$ and the dashed line represents $D=\sqrt{3/2}P_{2\otimes 2}$.}
\label{DiscordVsP22Scatter}
\end{figure}

Next, we consider $X$-states for which the analytic expression of quantum discord in a given computational basis is known \cite{Luo,Chen, Cheng}. The general form of the $X$-state is 
\begin{align}\label{Xstate}
\rho=
\left[\begin{array}{cccc}
\rho_{11} & 0 & 0 & \rho_{14}\\
0 & \rho_{22} & \rho_{23}& 0\\
0 & \rho_{32} & \rho_{33} & 0\\
\rho_{41} & 0 & 0 & \rho_{44}
\end{array}\right],
\end{align}
where $\rho_{11}+\rho_{22}+\rho_{33}+\rho_{44}=1$ and $\rho_{ij}=\rho_{ji}^*$ with $i,j=1,2,3,4$. Quantum discord for the $X$-state is \cite{Cheng} 
\begin{align}
D=\text{min}(Q_1,Q_2),\label{discord-x}
\end{align}
where $Q_1=\sum_{i=1}^4 \lambda_i \text{log}_2\lambda_i+H(\rho_{11}+\rho_{33})+H(\tau)$ and $Q_2=\sum_{i=1}^4 \lambda_i \text{log}_2\lambda_i+\sum_i \rho_{ii}\text{log}_2(\rho_{ii})$. Here $\lambda_i$'s are the eigenvalues of $X$-state, $H(x)=-x\text{log}_2 x-(1-x)\text{log}_2 (1-x)$ and $\tau=\frac{1}{2}(1+\sqrt{[1-2(\rho_{33}+\rho_{44})^2+4(|\rho_{14}|+|\rho_{23}|)^2})$. The intrinsic degree of coherence of the $X-$state can be shown to be 
\begin{align}
P_{2\otimes 2}=\sqrt{\frac{4}{3}}\left(\rho_{11}^2 \right. &+\rho_{22}^2 +\rho_{33}^2+\rho_{44}^2 \nonumber\\
& \left. +2(|\rho_{14}|^2+|\rho_{23}|^2)-\frac{1}{4}\right)^{1/2}.\label{P2-x}
\end{align}
Using Eqs.~(\ref{discord-x}) and (\ref{P2-x}), we generate a scatter plot with a set of 2 $\times 10^7$ randomly chosen $X$-states (see Fig.~\ref{DiscordVsP22Scatter}). The continuous line represents $D=P_{2\otimes 2}$, and the dashed line is $D=\sqrt{3/2}P_{2\otimes 2}$, which is the upper bound on discord for a given $P_{2\otimes 2}$. From the scatter plot, we find that the $X$-states follow a tighter upper bound, given by $D\leq P_{2\otimes 2}$. Furthermore, it can be seen that this tighter bound for $X$-states is saturable, that is, $D=P_{2\otimes 2}$ at least for $P_{2\otimes 2}=0, 1/3$ and 1. For example, the state $\rho_1=\ket{\psi^{\pm}}\bra{\psi^{\pm}}$, where $\ket{\psi^{\pm}}=\frac{1}{\sqrt{2}}(\ket{01}\pm\ket{10})$ saturates the bound with  $D=P_{2\otimes 2}=1$, while the states $\rho_2=\frac{1}{3}[\ket{00}\bra{00}+\ket{\psi^{\pm}}\bra{\psi^{\pm}}+\ket{11}\bra{11}]$ and $\rho_3=\frac{1}{3}[\ket{00}\bra{00}+\ket{\phi^{\pm}}\bra{\phi^{\pm}}+\ket{11}\bra{11}]$ yields $D=P_{2\otimes 2}=1/3$.

\section{Concurrence and intrinsic degree of coherence of two-qubit state}\label{Concurrence}

Concurrence is the most widely used measure of quantum entanglement \cite{wootters1998prl}. For a two-qubit state $\rho$, concurrence is given by
$C(\rho)=\text{max}(0,\lambda_1-\lambda_2-\lambda_3-\lambda_4),$
where $\lambda$'s are in descending order and are the square roots of the eigenvalues of matrix $\rho(\sigma_y\otimes \sigma_y)\rho^{*}(\sigma_y\otimes \sigma_y)$ with $\sigma_y$ being the Pauli matrix. A non-zero value of concurrence implies non-zero entanglement of the two-qubit state. It has been shown that concurrence of a two qubit state satisfies the following inequality \cite{Mintert}:
\begin{align}
C(\rho)\leq \sqrt{2[1-\text{Tr}(\rho_A^2)]},
\end{align}
where $\rho_A=\text{Tr}_B(\rho)$ is the reduced density matrix of subsystem $A$. The equality holds if and only if the state $\rho$ is pure. Using Eq.~(\ref{IDoCSubSystem}), we write the above inequality as
\begin{align}\label{ConcAndP2}
C(\rho)\leq \sqrt{1-(P_2^A)^2},
\end{align}
where $P_2^A$ is the intrinsic degree of coherence of subsystem $A$. Similarly, one can express $C(\rho)$ in terms of $P_2^B$ as $C(\rho)\leq \sqrt{1-(P_2^B)^2}$. Using this and the above inequality, we obtain
\begin{align}
[C(\rho)]^2\leq 1-\frac{(P_2^A)^2+(P_2^B)^2}{2}.
\end{align}
The above inequality reflects a well-known fact that for a maximally-entangled two-qubit state, that is, for $C(\rho)=1$, the individual qubits are completely mixed, that is, $P_2^A=P_2^B=0$. Now, it is known that concurrence $C(\rho)$ satisfies the following inequality  \cite{Mintert}
\begin{align}
[C(\rho)]^2\geq 2[\text{Tr}(\rho^2)-\text{Tr}(\rho_A^2)].
\end{align} 
Thus, using Eqs. (\ref{IDOC}) and (\ref{IDoCSubSystem}), we write the above inequality as
\begin{align}\label{lowerBoundConc}
[C(\rho)]^2\geq \frac{3}{2}P_{2\otimes 2}^2-\frac{1}{2}-(P_2^A)^2.
\end{align}
We find that the intrinsic degree of coherence of the two-qubit state together with that of the individual qubits decide the range of values that the concurrence of the two-qubit state can take. Although a closed-form analytic expression for concurrence of a general two-qubit state is available, the motivation behind the derivation of the above bound is to understand how the entanglement of a two-qubit state gets dictated by the intrinsic degree of coherence of the two-particle system and that of the individual subsystems. We note that for Bell states the above inequality gets saturated and becomes the inequality given in Eq.~(\ref{ConcAndP2}). We further note that the inequality given in Eq.~(\ref{lowerBoundConc}) provides an entanglement criterion, which says that if a two-qubit state $\rho$ satisfies $3P_{2 \otimes 2}^2-1 >2 (P_2^A)^2$ then it must be entangled.

\section{Measuring the intrinsic degree of coherence of two-qubit states}\label{Experiment}

\begin{figure}[b !]
\centering
\includegraphics[scale=0.9]{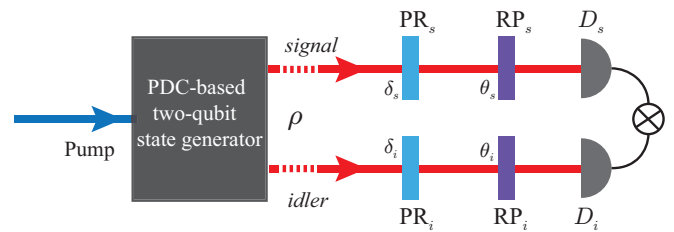}
\caption{Schematic of the proposed experimental setup for measuring the intrinsic degree of coherence of a two-qubit state. $PR_s$ and $PR_i$ are phase retarders; $RP_s$ and $RP_i$ are rotation plates; and $D_s$ and $D_i$ are photon detectors in a coincidence-counting setup. }
\label{ExperimentalSetup}
\end{figure}

In this section, we propose an experimental scheme for measuring the intrinsic degree of coherence $P_{2\otimes 2}$ of a generic polarization two-qubit state produced by utilizing the nonlinear optical process of  parametric down-conversion (PDC). In PDC, a pump photon splits into two entangled photons called the signal and idler photons \cite{Boyd}. The schematic of the experimental scheme is depicted in Fig.~\ref{ExperimentalSetup}. In the figure, the PDC-based two-qubit state generator produces an arbitrary two-qubit state in the polarization basis. Such two-qubit state generators can be realized by involving multiple PDC crystals. One such example is the two-qubit state generator proposed in Ref.~\cite{Kulkarni} using four PDC cyrstals. In Fig.~\ref{ExperimentalSetup}, the signal (idler) photon passes through  the phase retarder $PR_s(PR_i)$ and the rotation plate $RP_s(RP_i)$ and is detected by the detector $D_S (D_i)$.  The rotation plate $RP_s(RP_i)$ rotates the polarization of the signal(idler) photon by angle $\theta_s(\theta_i)$ while the phase retarder $PR_s(PR_i)$ introduces a phase difference $\delta_s(\delta_i)$ between the horizontal and vertical polarization components of the signal (idler) field. We note that the phase retarders and rotation plates can be realized using ordinary waveplates \cite{Simon1989PhysLettA}. We now consider the most general two-qubit state $\rho$ and write it as \cite{byrd2003pra}
\begin{align}\label{densitymatrixSU4}
\rho=\frac{1}{4}\left(I+\sqrt{6}\sum_{j=1}^{15} r_j \Lambda_j\right).
\end{align}
Here $r_j$'s are the analogs of Stokes parameters and $\Lambda_j$'s are the $4\times 4$ generalized Gellmann matrices, which are the generators of the group SU(4) \cite{kimura2003pla}. Using Eq.~(\ref{IDOC}), we write the degree of coherence $P_{2\otimes 2}$  in terms of the Stokes parameters $r_j$ as \cite{Patoary}
\begin{align}\label{P22rj}
P_{2\otimes 2}=\sqrt{\sum_{j=1}^{15} |r_j|^2}.
\end{align}
Thus, we have that $P_{2\otimes 2}$ can be measured experimentally by measuring the Stokes parameters. We now outline as to how this measurement could be performed. The state of the two qubits just before the detectors is given by $\tilde{\rho}=U\rho U^\dagger$, where $U=(U_s^\theta U_s^\delta)\otimes (U_i^\theta U_i^\delta)$ with
\begin{align}
U_k^\theta=
\left(\begin{array}{cc}
\cos\theta_k & \sin\theta_k\\
-\sin\theta_k & \cos\theta_k
\end{array}\right), \ \  U_k^\delta=\left(\begin{array}{cc}
1 & 0\\
0 & e^{i\delta_k}
\end{array}\right),
\end{align}
and $k=s,i$. The signal and idler photons are detected in coincidence at $D_s$ and $D_i$ in the horizontal polarization directions. Therefore, the coincidence detection probability $M$ can be written as $M=\langle HH| U\rho U^\dagger|HH\rangle$, evaluating which, we get
\begin{align}\label{coincidenceintensity}
&M=\cos\theta_s\cos\theta_i\left[\cos\theta_s(\rho_{11} \cos\theta_i+\rho_{21}e^{-i\delta_i}\sin\theta_i )\right. \nonumber\\
&\left.+\sin\theta_s(\rho_{31}e^{-i\delta_s}\cos\theta_i+\rho_{41}e^{-(\delta_s+\delta_i)}\sin\theta_i)\right]\nonumber\\
&+e^{i\delta_i}\cos\theta_s\sin\theta_i\left[\cos\theta_s(\rho_{12} \cos\theta_i+\rho_{22}e^{-i\delta_i}\sin\theta_i )\right. \nonumber\\
&\left.+\sin\theta_s(\rho_{32}e^{-i\delta_s}\cos\theta_i+\rho_{42}e^{-(\delta_s+\delta_i)}\sin\theta_i)\right]\nonumber\\
&+e^{i\delta_s}\sin\theta_s\cos\theta_i\left[\cos\theta_s(\rho_{13} \cos\theta_i+\rho_{23}e^{-i\delta_i}\sin\theta_i )\right. \nonumber\\
&\left.+\sin\theta_s(\rho_{33}e^{-i\delta_s}\cos\theta_i+\rho_{43}e^{-(\delta_s+\delta_i)}\sin\theta_i)\right]\nonumber\\
&+e^{i(\delta_s+\delta_i)}\sin\theta_s\sin\theta_i\left[\cos\theta_s(\rho_{14} \cos\theta_i\right. \nonumber\\
&\left.+\rho_{24}e^{-i\delta_i}\sin\theta_i )+\sin\theta_s(\rho_{34}e^{-i\delta_s}\cos\theta_i\right.\nonumber\\
&\left.+\rho_{44}e^{-(\delta_s+\delta_i)}\sin\theta_i)\right].
\end{align}
We find that by measuring $M$ at various settings of parameters $\theta_s$, $\theta_i$, $\delta_s$, and $\delta_i$, one can calculate the Stokes parameters. Table \ref{TableExp} shows a convenient set of 16 measurement settings of these parameters. In the table, we have denoted the coincidence probabilities by $M_i$, with $i=1,2, \cdots 16$ being the measurement index. Using Eqs.~(\ref{densitymatrixSU4}) through (\ref{coincidenceintensity}), one can show that the Stokes parameters are related to the 16 coincidence probabilities as
\begin{subequations} 
\begin{eqnarray}
&r_1=\frac{-1}{2}(M_1+M_2-2M_5),\\
&r_2=\frac{-1}{2}(M_1+M_3-2M_9),\\
&r_3=\frac{1}{2}[M_{12}-M_{11}+M_{10}-M_9+M_8 \nonumber\\
&~~~~~~~-M_7+M_6-M_5+2(M_{13}-M_{16})],\\
&r_4=\frac{1}{2}[2(M_{13}+M_{16})-(\sum_{j=5}^{12}M_j-\sum_{j=1}^4 M_j)],~~~~\\
&r_5=\frac{-1}{2}(M_2+M_4-2M_{11}),\\
&r_6=\frac{-1}{2}(M_3+M_4-2M_7),\\
&r_7=\frac{1}{2}(M_1+M_2-2M_6),\\
&r_8=\frac{1}{2}(M_1+M_3-2M_{10}),\\
&r_9=\frac{-1}{2}[2(M_{14}+M_{15})-(\sum_{j=5}^{12}M_j-\sum_{j=1}^4 M_j)],~~~\\
&r_{10}=\frac{-1}{2}[-M_{12}+M_{11}-M_{10}+M_9+M_8 \nonumber\\
&~~~~~~~-M_7+M_6-M_5+2(M_{14}-M_{15})],\\
&r_{11}=\frac{1}{2}(M_2+M_4-2M_{12}),\\
&r_{12}=\frac{1}{2}(M_3+M_4-2M_{8}),\\
&r_{13}=\frac{1}{2}(M_1-M_2),\\
&r_{14}=\frac{1}{2\sqrt{3}}(M_1+M_2-2M_{3}),\\
&r_{15}=\frac{1}{2\sqrt{6}}(M_1+M_2+M_3-3M_4).
\end{eqnarray}    
\end{subequations}\label{stokesparameters}
Therefore, in order to measure $P_{2\otimes 2}$, one needs to first measure $M_1$ through $M_{16}$ experimentally, then evaluate the Stokes parametes using the above equation, and finally substitute these Stokes parameters in Eq.~(\ref{P22rj}). We note that while the above state reconstruction procedure is sufficient in principle, additional post-processing may be required for a reliable reconstruction in the presence of experimental noise \cite{James2001PRA}. 
\begin{table}[h!]
\begin{center}
\begin{tabular}{ |c|c|c|c|c| } 
 \hline
 $~\theta_s~$ & $~\theta_i~$ & $~\delta_s~$ & $~\delta_i~$   & $~M~$ (Coincidence counts) \\ 
 \hline
 \hline
 0 & 0 & 0  &  0  & $M_1$ \\ 
 \hline
 0 & $\pi/2$ & 0  &  0  & $M_2$ \\ 
 \hline
 $\pi/2$ & 0 & 0  &  0  & $M_3$ \\ 
 \hline
 $\pi/2$ & $\pi/2$ & 0  &  0  & $M_4$ \\ 
 \hline
 0 & $\pi/4$ & 0  &  0  & $M_5$ \\ 
 \hline
 0 & $\pi/4$ & 0  &  $\pi/2$  & $M_6$\\ 
 \hline
 $\pi/2$ & $\pi/4$ & 0  &  0  & $M_7$ \\ 
 \hline
 $\pi/2$ & $\pi/4$ & 0  &  $\pi/2$  & $M_8$ \\ 
 \hline
 $\pi/4$ & 0 & 0  &  0  & $M_9$ \\ 
 \hline
 $\pi/4$ & 0 & $\pi/2$  &  0  & $M_{10}$ \\ 
 \hline
 $\pi/4$ & $\pi/2$ & 0  &  0  & $M_{11}$ \\ 
 \hline
 $\pi/4$ & $\pi/2$ & $\pi/2$  &  0  & $M_{12}$ \\ 
 \hline
 $\pi/4$ & $\pi/4$ & 0  &  0  & $M_{13}$ \\ 
 \hline
 $\pi/4$ & $\pi/4$ & $\pi/2$  &  0  & $M_{14}$ \\ 
 \hline
 $\pi/4$ & $\pi/4$ & 0  &  $\pi/2$  & $M_{15}$ \\ 
 \hline
 $\pi/4$ & $\pi/4$ & $\pi/2$  &  $\pi/2$  & $M_{16}$ \\ 
 \hline
\end{tabular}
\caption{Coincidence probabilities $M_i$, with $i=1,2, \cdots 16$, at 16 different settings of the parameters  $\theta_s$, $\theta_i$, $\delta_s$, and $\delta_i$.}
\label{TableExp}
\end{center}
\end{table}

\section{Transfer of intrinsic degree of coherence in PDC}\label{SPDCDiscord}

In recent years, transfer of the coherence properties of the pump photon to the signal and idler photons has been studied in various different contexts \cite{Jha, Kulkarni, kulkarni2017pra}. In this section, we investigate this transfer in terms of the intrinsic degree of coherence. For conceptual clarity, we restrict our analysis to polarization two-qubit states. We take the pump field to be partially polarized and represent it by a $2\times 2$ density matrix $\rho_{\text{pump}}$ in the polarization basis. We represent the generated two-qubit state by a $4\times 4$ density matrix $\rho_{\rm si}$. The eigenvalues of $\rho_{\text{pump}}$ are denoted by $\epsilon_1$ and $\epsilon_2$ and those of $\rho_{\rm si}$ are denoted by $\lambda_1,\lambda_2,\lambda_3$, and $\lambda_4$. It can be very easily shown that $\epsilon_1=(1+P_2^{\text{pump}})/2,\epsilon_2=(1-P_2^{\text{pump}})/2$ \cite{Mandel,Kulkarni}, where $P_2^{\text{pump}}$ is the intrinsic degree of coherence of the pump field and is given by $P_2^{\text{pump}}=\sqrt{2 \text{Tr}(\rho_{\text{pump}}^2)-1}=\sqrt{2(\epsilon_1^2+\epsilon_2^2)-1}$. The intrinsic degree of coherence of the generated two-qubit state $\rho_{\rm si}$ is given by $P_{2\otimes 2}^{\text{si}}=\sqrt{[4 \text{Tr}(\rho_{\rm si}^2)-1]/3}=\sqrt{\left(4\sum_i \lambda_i^2-1 \right)/3}$. We assume that the two-photon state generation process is trace-preserving and entropy non-decreasing. Under these assumptions it can be shown based on majorization  \cite{Grahame,Kulkarni} that $
\lambda_1^2+\lambda_2^2+\lambda_3^2+\lambda_4^2 \leq \epsilon_1^2+\epsilon_2^2$. Using this majorization relation, we obtain
\begin{align}\label{P22si}
P_{2\otimes 2}^{\text{si}}\leq \sqrt{\frac{1+2(P_2^{\text{pump}})^2}{3}}.
\end{align}
The equality holds in situations in which the two-qubit state generation process is unitary. The two-qubit generation process becomes non-unitary in the presence of scattering or decohering channels. Thus, we find that the intrinsic degree of coherence of the pump field puts an upper bound on the intrinsic degree of coherence of the down-converted two-qubit state. For unitary generation processes, $P_2^{\text{pump}}$ fixes the value of $P_{2\otimes 2}^{\text{si}}$ through the relation  $P_{2\otimes 2}^{\text{si}}= \sqrt{\frac{1+2(P_2^{\text{pump}})^2}{3}}$. The intrinsic degree of coherence of the two-qubit state $P_{2\otimes 2}^{\text{si}}$ becomes unity when the two-qubit generation process in unitary and when the pump is completely polarized, that is $P_2^{\text{pump}}=1$.

\section{Summary}\label{Summary}

In this article, we have defined the intrinsic degree of coherence of two-qubit states and have demonstrated its usefulness in quantifying the quantum correlations and entanglement of a two-particle system. We have shown that the intrinsic degree of coherence of a two-qubit state puts an upper bound not only on the violations of Bell inequalities but also on quantum discord. Furthermore, the range of values that the concurrence of a two-qubit state can take is shown to be decided by the intrinsic degree of coherence of the two-qubit state together with that of the individual qubits. Finally, in the context of PDC, we have proposed an experimental scheme for measuring the intrinsic degree of coherence of two-qubit states and have studied how it depends on the intrinsic degree of coherence of the pump.

\section*{Acknowledgments}
We acknowledge financial support through the research
grant no. EMR/2015/001931 from the Science and Engineering
Research Board (SERB), Department of Science
and Technology, Government of India and through
the research grant no. DST/ICPS/QuST/Theme-
1/2019 from the Department of Science and Technology,
Government of India. NM acknowledges IIT Kanpur for postdoctoral fellowship. \\

\section*{Disclosures}
The authors declare no conflicts of interest.
%

\end{document}